# Super-resolution reconstruction of cytoskeleton image based on A-net deep learning network

Qian Chen[1,#], Haoxin Bai [2,#], Bingchen Che[2], Tianyun Zhao[1,*], Ce Zhang[2,*], Kaige Wang[2], Jintao Bai[2], Wei Zhao[2,*]

*1. School of Automation, Northwestern Polytechnical University, Xi'an 710129, China*

*2. State Key Laboratory of Photon-Technology in Western China Energy, International Collaborative Center on Photoelectric Technology and Nano Functional Materials, Institute of Photonics & Photon Technology, Northwestern University, Xi'an 710127, China*

# These authors have equal contributions to this work

*\* Correspondence: zhaoty@nwpu.edu.cn; zhangce@nwu.edu.cn; zwbayern@nwu.edu.cn*

**Abstract**：To date, live-cell imaging at the nanometer scale remains challenging. Even though super-resolution microscopy methods have enabled visualization of subcellular structures below the optical resolution limit, the spatial resolution is still far from enough for the structural reconstruction of biomolecules in vivo (i.e., ~ 24 nm thickness of microtubule fiber). In this study, we proposed an A-net network and showed that the resolution of cytoskeleton images captured by a confocal microscope can be significantly improved by combining the A-net deep learning network with the DWDC algorithm based on degradation model. Utilizing the DWDC algorithm to construct new datasets and taking advantage of A-net neural network's features (i.e., considerably fewer layers), we successfully removed the noise and flocculent structures, which originally interfere with the cellular structure in the raw image, and improved the spatial resolution by 10 times using relatively small dataset. We, therefore, conclude that the proposed algorithm that combines A-net neural network with the DWDC method is a suitable and universal approach for exacting structural details of biomolecules, cells and organs from low-resolution images.

## 1. Introduction

The micro-scale organizations and nano-scale biomolecular structures play essential roles in life machinery, e.g., the nano-pores controls transportation [1] and cytoskeleton behaves as mechano-sensor [2]. To understand the underlying mechanism of cellular behavior, it is important to monitor the dynamics of biomolecules at resolution of tens of nanometers, e.g., the ~ 50 nm persistence length of DNA [3] and ~ 24 nm thickness of microtubule fiber [4]. Imaging platforms, which are reported to achieve such resolution, include Transmission Electron Microscope (TEM, 300 nm) [5], Scanning Electron Microscope (SEM, 200 nm) [6], Cryogenic electron microscopy (Cryo-EM, 200 nm) [7] and Stimulated Emission

Depletion (STED, 20 nm) microscopy [8], etc. But, TEM, SEM and Cyto-EM are not suitable for live-cell imaging and monitoring molecular dynamics *in vivo* [9]. STED microscopy is a promising technique. Its application is, however, hindered by the presence of specific fluorophores [10], the overly complex operational procedures [11], and the high cost [12]. Considering the fact that there exist large quantities of image data in various databases [13], and most labs are only equipped with commonplace inverted microscopes with submicron resolution and high noise level, it is critical to develop a numerical approach, which can exact molecular information from poor quality images.

Currently, reported image process algorithms can be categorized as traditional [14, 15], and deep-learning image process algorithms[16-18]. The latter has become a focus of image processing community, and many algorithms have been developed. For instance, Super-Resolution Convolutional Neural Network (SRCNN) [19] is an end-to-end network developed based on sparse coding to get a sharper edge and higher resolution of images. The shortcomings of SRCNN include the sacrifice of processing speed to realize an acceptable restoration quality. Fast Super-Resolution Convolutional Neural Network (FSRCNN) [20] is an update of SRCNN. It gets a big speed boost, while simultaneously lose details as a result of over smooth. Super-Resolution Generative Adversarial Network (SRGAN) [21] optimizes loss function to obtain high PSNR (Peak SNR) and enhance the restored image's sense of reality. Visually, the restored images show better reality, however, PSNR of images is reduced. Besides, these algorithms primarily aim to enrich the pixel information of the image, not to improve the intrinsic optical resolution of image.

We herein propose an A-net network which is realized by improving the structure of U-net network. Accompanied with a traditional degradation model to process label images, the details of microtubule network captured by confocal microscope can be extracted with higher resolution and SNR. In brief, raw images were firstly processed by threshold denoising and a three-dimensional Gaussian interpolation. We, then, obtained the corresponding label images using DWDC [22] method, which combines discrete wavelet and Lucy-Richardson deconvolution [23]. The pairs of original images with the corresponding label images served as our own datasets, relying on which the A-net network was trained. Finally, the test images were processed according to the A-net network. It is demonstrated that our method can effectively remove noise and flocculent structures in the raw images, resulting in increased resolution by ~10 times.

## 2. Related works

As the purpose of this paper is to explore super-resolution algorithms based on neural networks, we briefly review the existing algorithms for improving image resolution, followed by detailed introduction on the super-resolution algorithms based on deep learning.

*2.1 Traditional methods*

Traditional image processing algorithms mainly rely on basic digital image processing techniques. Generally, there are three categories: interpolation-based algorithms [24-26], degenerate model-based algorithms [27-29] and learning-based algorithms [18, 30-32].

Interpolation-based algorithms [33, 34] use the original pixel information of the low-resolution image to "guess" the subpixel information of image based on interpolation. It can effectively upgrade the low-resolution image to high resolution image with more pixels. Nevertheless, in practical applications, the interpolation algorithms can only improve the image details in a very limited way.

Degenerate model-based algorithms [35, 36] focus on establishing an observation model for the acquisition process of images, and then realize super-resolution reconstruction by solving the inverse problem of the observation model. The observation model describes the process of obtaining the low-resolution observation image from the high-resolution image by the imaging system, as shown in formula (1):

$$L = H * f + N \tag{1}$$

where $L$ is the low-resolution image, $H$ is the high-resolution image, $f$ is a transformation function (i.e. the point spread function in optical system), $N$ is noise. This method restores the actual information of object with a higher resolution, based on the estimation of $f$. Commonly, this type of super-resolution algorithm includes iterative back projection (IBP) [36], projection of convex set (POCS) [28], maximum posterior probability (MAP) [37] and Bayesian analysis [38, 39] methods etc. These methods aim to get better visual quality of images and restore object details. However, they also suffer a series of problems, e.g. the processing speed is generally slow and may lead to spurious images.

Learning-based algorithms [40, 41] aim to build a mapping between the low-resolution image and the corresponding high-resolution image by prior training and learning from the dataset. Learning-based algorithms are mainly realized by machine learning. There are several commonly used machine learning methods, including neighborhood embedding [42], support vector regression [43, 44], manifold learning [45], and sparse representation [46, 47] etc. Learning-based algorithms are limited by several disadvantages including the need to manually optimize parameters and the lack of end-to-end training, which leads to poor algorithm applicability.

*2.2 Deep-learning-based algorithm*

In recent years, various deep learning-based super-resolution algorithms have been developed. Dong et al [19] firstly applied deep neural network to super-resolution processing. They proposed SRCNN to learn the end-to-end mapping between low-resolution images and corresponding high-resolution images. A three-layer convolutional neural network is combined with bilinear interpolation and nonlinear mapping to form the SRCNN algorithm. SRCNN algorithm automatically optimizes all parameters by learning from the input training set, and can therefore, reach an average PSNR value of 30.09 dB with a runtime of 0.18 sec per image. Dong et al [20] developed FSRCNN based on SRCNN. A deconvolution layer was used at the end of the network to enlarge the image size, which can save time by eliminating the pre-training phase. The network also replaces convolution kernels in SRCNN with smaller convolution kernels and share convolution layers in order to reduce the computation. These improvements help the network reduce the calculation parameters and speed up the processing. As a consequence, FSRCNN has very fast processing speed without the loss of restoration quality and achieves an average PSNR of 32.87 dB with a processing speed of 24.7 fps.

Kim et al [48] extended the network layers to 20 layers based on SRCNN and introduced residual structure [49] into the network, i.e. Very Deep Convolutional Networks (VDSR). The deep network layer has a larger receptive field, and more information can be learned with better accuracy. The VDSR network uses the residual learning method to limit the gradient to a certain range, which can speed up the convergence process. As compared to SRCNN, the VDSR network realized a higher accuracy, a faster convergence, and resolution of folds higher. In their investigation, VDSR achieves an average PSNR of 37.53 dB.

In the aforementioned methods, over smooth of image is inevitable and could lead to spurious image. Ledig et al [21] proposed a generative adversarial network (SRGAN) which is developed on the basis of Generative Adversarial Network(GAN) [50] to solve the problem for super-resolution processing and recovering finer texture structures. The Generate network generates high-resolution prediction images from low-resolution original images, and the Discriminate network determine whether the prediction image is consistent to the corresponding label image. Although the PSNR values were not apparently improved, the details of the image have been enhanced to super-resolution level. It should be noted, again, they aim to enrich the pixel information of the image, e.g. from 512×512 pixels to 1024×1024 pixels, not to improve the intrinsic optical resolution of image, e.g. from 300 nm to 100 nm.

## 3. Algorithm

To improve the resolution of poor quality image intrinsically, i.e. extract the real structure from a blurred and noisy image, we proposed A-net by improving the structure of U-net network, and proposed an approach, combining traditional image preprocessing algorithms based on degenerate model with A-net network [51]. The overall framework of this algorithm is schematically shown in Fig. 1. For the A-net deep learning network, the image pairs of original and label images are required to construct train datasets for deep neural network. Because of the scarcity and particularity of biological images, we need to build our own biological microtubule image dataset (i.e., SR_MUI dataset), which were obtained using DWDC[22] method that constituted of a series of pre-processing methods, discrete wavelet method, Lucy-Richardson deconvolution method and post-processing methods. The training dataset is input into the network so that A-net network can learn the mapping relationship between low-resolution images and high-resolution label images. Then the test dataset is input into A-net network for prediction, and the super-resolution images are obtained.

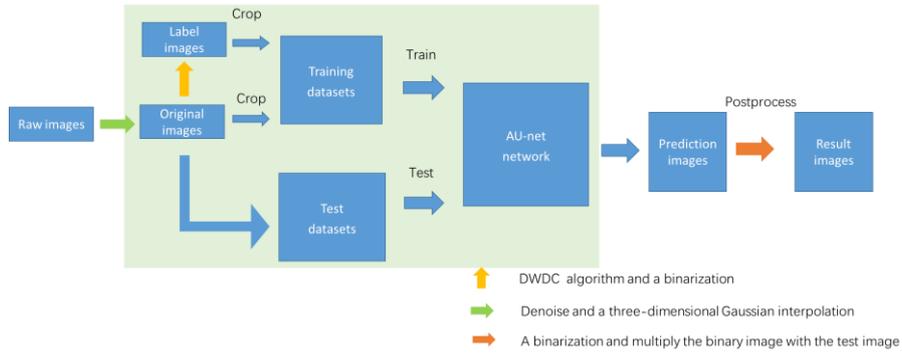

Fig. 1. Overall framework of the algorithm. A series of preprocessing methods are used to obtain the original image and the label image on the raw image. Denoise and a three-dimensional Gaussian interpolation is performed on the raw confocal images to get the original image, as shown by the green arrow. Then, the DWDC algorithm and a binarization is used to get high-resolution label images (shown by the yellow arrow). The original image and the corresponding label image are composed of image pairs and cropped into 512×512 size to construct the biological microtubule image dataset referred to as SR_MUI dataset. A-net network trains the parameters in the network through the image of the training dataset, and then get the corresponding prediction results. In the postprocessing, as shown by the orange arrow, we apply binarization on the prediction image to get a shrinkage outline of the microtubule structures. Subsequently, the binary image is multiplied with the test image to get the result image.

*3.1 Raw images and processing targets*

The raw images to be processed in this paper are confocal fluorescent images of 3T3 fibroblast microtubule labeled by tubulin fluorescent dye, which is excited at 640 nm and emitted around 674 nm (Fig. 2). The raw images were captured by a commercial confocal microscope (Nikon A1 LFOV) using the objective lens of Olympus 100X NA1.4 oil immersion lens. Each of the raw confocal image is in 16-bit TIF format with a size of 512×512 pixels. The pixel interval is 0.25 μm.

We observe that the filament-like microtubule structures are widely present in cells, characterized by poor SNR, insufficient spatial resolution and a lot of noise information. In the investigation, it is our goal to develop a universal algorithm to obtain super-resolution images of microtubules from such low-resolution image.

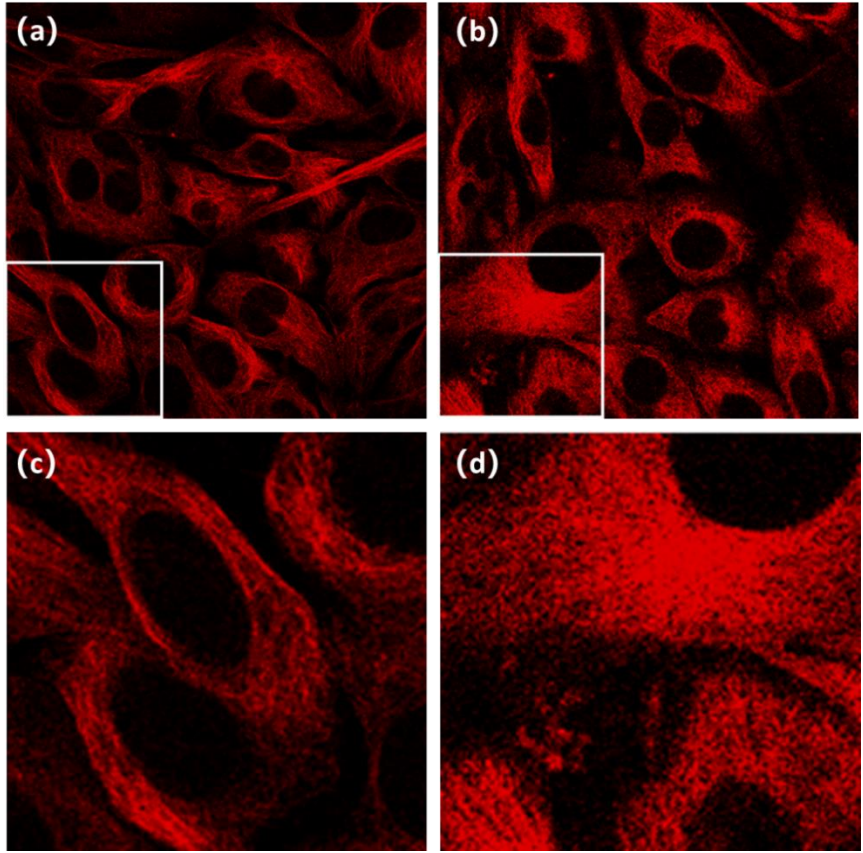

Fig. 2. Raw images captured by confocal microscope. The raw image are 3t3 cell microtubule images captured by a commercial confocal microscope (Nikon A1 LFOV) with the objective lens of Olympus 100X NA1.4 oil immersion lens. (a) and (b) show the image of different structures and shapes, (c) and (d) shows details of local regions in the white boxes of (a) and (b) separately. The raw images are all of 512×512 size.

*3.2 Preprocessing*

In order to get better image SNR, threshold denoising is used to reduce image noise. Since the pixel interval in the raw image is 0.25 μm, which affects the image resolution, a 3-dimensional Gaussian interpolation is performed twice, to extend the image size from 512×512 to 2048×2048, leading to reduced pixel interval of 63 nm. Then the DWDC algorithm is used to get high-resolution label images [22]. In addition, in order to extract the sketch of microtubule structures and prevent the details information to be immersed by background, we applied binarization accompanied with threshold processing to generate label images.

In the process, the expansion brings heavy burden to the server and network for computing. To improve the efficiency of training, a series of preprocessing have been made, as diagramed in Fig. 3. On one hand, we convert the 16-bit TIF images to 8-bit TIF images by projecting data from 16-bit to 8-bit with approximately linear way. On the other hand, crop the image (2048×2048) into sub-images (512×512). Construct the SR_MUI dataset by forming image pairs of the sub-image of original images and the corresponding sub-image of label images.

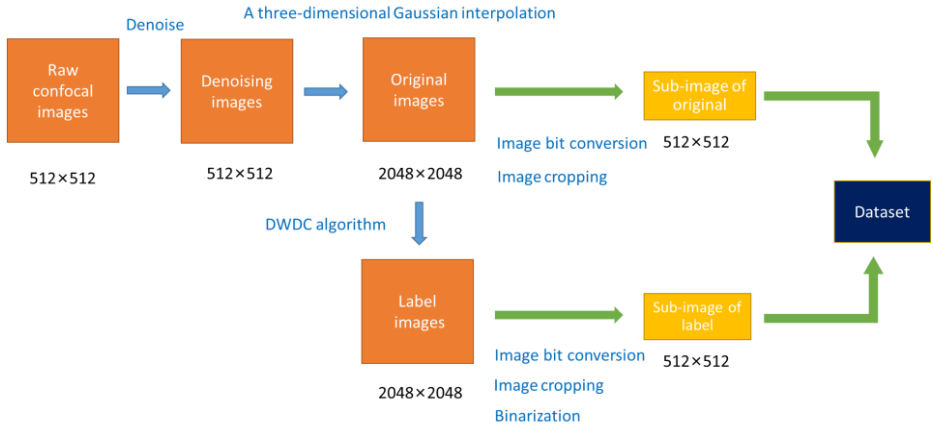

Fig. 3. Diagram of the pre-processing procedures. At the beginning, both threshold denoising and a three-dimensional Gaussian interpolation have been carried out on the raw confocal image. The image after these processing is adopted as original image for A-net network. Then, we use DWDC algorithm to get high-resolution label images. The label image is further binarized to prevent the network from learning additional feature information. Subsequently, both the original and label images are converted from 16-bit data to 8-bit, and cropped from 2048×2048 images into 512×512 sub-images, to reduce the load of A-net computation. Finally, the corresponding sub-image pairs form the SR_MUI dataset.

It is worth mentioning that since the image size of the training set is 512×512, when the test image is input into the A-net network, each of the test image needs to be divided into 16 sub-images of 512×512 pixels. These sub-images of test images are processed by the A-net network and the corresponding super-resolution prediction images are obtained.

*3.3 A-net network*

In this investigation, we focus on the filament-like microtubule structures, which can be approximate to a cluster or mesh of segments. Thus, U-net network which has been widely used in image segmentation was applied here for biostructure extraction and super-resolution processing. One of the most significant advantages of the U-net is that it does not require a big biological dataset. This is particularly important for us, since our dataset is relatively small, and there have been no established works or public datasets can realize our purpose.

The U-net network is composed of the encoder network and the decoder network with symmetric structures. In the encoder network, there are four convolution blocks for feature maps of different sizes. In the convolution block, there are two $3\times3$ convolutions in sequence, and followed by $2\times2$ max pooling. In the decoder network, there are also four deconvolution blocks corresponding to the encoder network. In the deconvolution block, there are two $3\times3$ convolutions in sequence, followed by a transposed convolution. The encoder network doubles the number of channels, reduces the sample size of the feature map by half. The decoder network doubles the size of the feature map and half the channel numbers. Therefore, the encoder-decoder network transforms the input image into small-size and multi-channel feature maps, and then decodes the feature map to output image with the same size. At the same time, we adopt skip-connection in the U-net network. This operation can connect feature maps in different sizes, which is helpful for gradient propagation and network convergence. All the convolutions in this neural network are followed by a Batch Normalization (BN) [52, 53] and a Rectified Linear Unit (ReLU) [54, 55] for faster training and preventing gradient vanishing problem.

Since the sizes of input image and output image of the U-net network are inconsistent, in order to make the output images of U-net network has the same size with the input image, we replace all valid convolution in the network with same convolution. The employment of same convolution makes the feature maps of the corresponding layers in encoding network and decoding network are exactly the same size. Thereafter, it is appropriate to directly copy the feature map of the encoding network to the decoding network, as shown in Fig. 4, and combine it with the feature map of the decoding network through the skip connection. This process avoids the crop operations in U-net network which can simplify the processing and reduce the image mismatching during cropping. Accordingly, the revised U-net network is named as A-net in this paper.

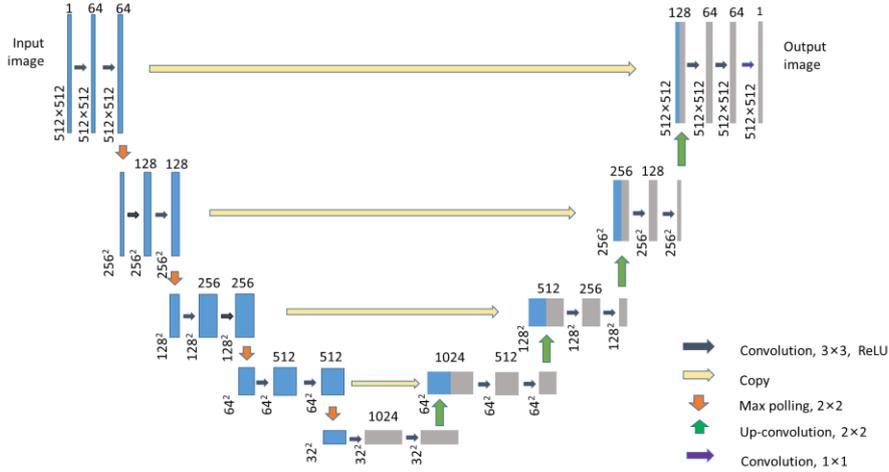

Fig. 4. A-net network architecture. In the network framework, the blue box denotes different feature maps in different layers. The corresponding channel numbers are provided on the top of the box. The white box denotes copied feature maps. Different colored arrows represent different operations which have been labelled on the right bottom of the figure.

The loss function of A-net is calculated by combining the cross entropy [56] loss function with a pixel-wise soft-max on the final feature map. The soft-max function [57] can be calculate as following

$$P_i(x) = \frac{\exp[a_i(x)]}{\sum_{j=1}^{M} \exp[a_j(x)]} \qquad (2)$$

where $P_i(x)$ denotes the approximated maximum-function, $i$ represents the category of pixels, $a_i(x)$ represents the activation function score of the category of pixel is $i$ with the pixel position $x \in \Omega$ and $\Omega \subset \mathbb{Z}^2$, M represents the number of classes, $a_j(x)$ represents the activation function score when the category of image pixel points is $j$, $\sum_{j=1}^{M} \exp[a_j(x)]$ represents the sum of all classes of activation functions. In conclusion, $P_i(x)$ is the classification result of pixel $x$ of class $M$, maximizing the most likely result while suppressing the probability of other categories. The sum of probabilities of all prediction categories is 1. For the $i$ that has the maximum activation $a_i(x)$, the responding $P_i(x) \approx 1$. While for all the other $i$, the responding $P_i(x) \approx 0$. Then, cross entropy penalizes $P_{g(x)}(x)$ for a deviation from 1 at every position by Eq. (3) [56]

$$E = \sum_{x \in \Omega} \omega(x) \log [P_{g(x)}(x)] \tag{3}$$

where $\omega \in \Omega$ with $\Omega \subset \mathbb{R}$ denotes a weight, $g(x)$ denotes the ground truth of each pixel. The purpose of setting $\omega$ is to give higher weights to pixels in the image that are close to the boundary points. The weight graph is calculated in advance with ground truth of each pixel in label images, in order to let network to learn to distinguish smaller boundaries.

### 3.4 Postprocessing

The A-net network predicts the input sub-images (512×512 pixels) of the test image and obtains corresponding predicted sub-images (512×512 pixels), which are subsequently assembled as prediction image (2048×2048 pixels). Then, we apply binarization on the prediction image in order to get a shrinkage outline of the microtubule structures. Subsequently, the binary image is multiplied with the test image to get the result image (will be shown in the Result section).

## 4. Experiments

### 4.1 SR_MUI dataset

We constructed a biological microtubule image dataset, i.e. SR_MUI dataset, based on 3t3 cell images. The raw confocal image, the original image, and the high-resolution label images are shown in Fig. 5.

In SR_MUI dataset, there are 200 image pairs for training and 50 images for testing. A preview of SR_MUI dataset is shown in Fig. 6. It can be seen, the label images clearly extract the sketch of the microtubule structures from the noisy and blur raw images.

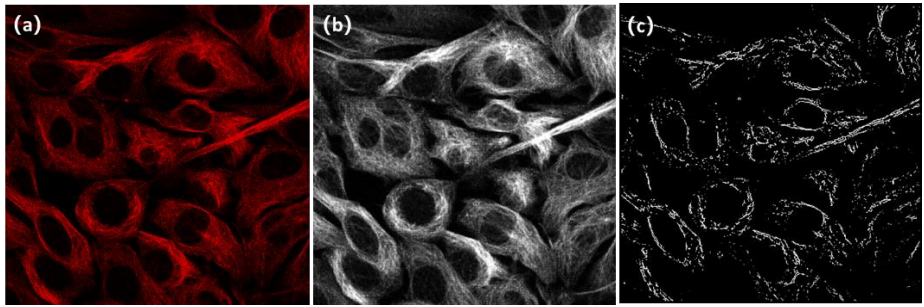

Fig. 5. Images during the production of SR_MUI dataset. (a) The raw 3t3 cell images captured by confocal microscope (Nikon A1 LFOV), the image size is 512×512, (b) the image obtained by threshold denoising algorithm and three-dimensional Gaussian interpolation algorithm from (a), the image size is 2048×2048, (c) shows the high-resolution label images obtained by DWDC algorithm and a binarization from (a), the image size is 2048×2048.

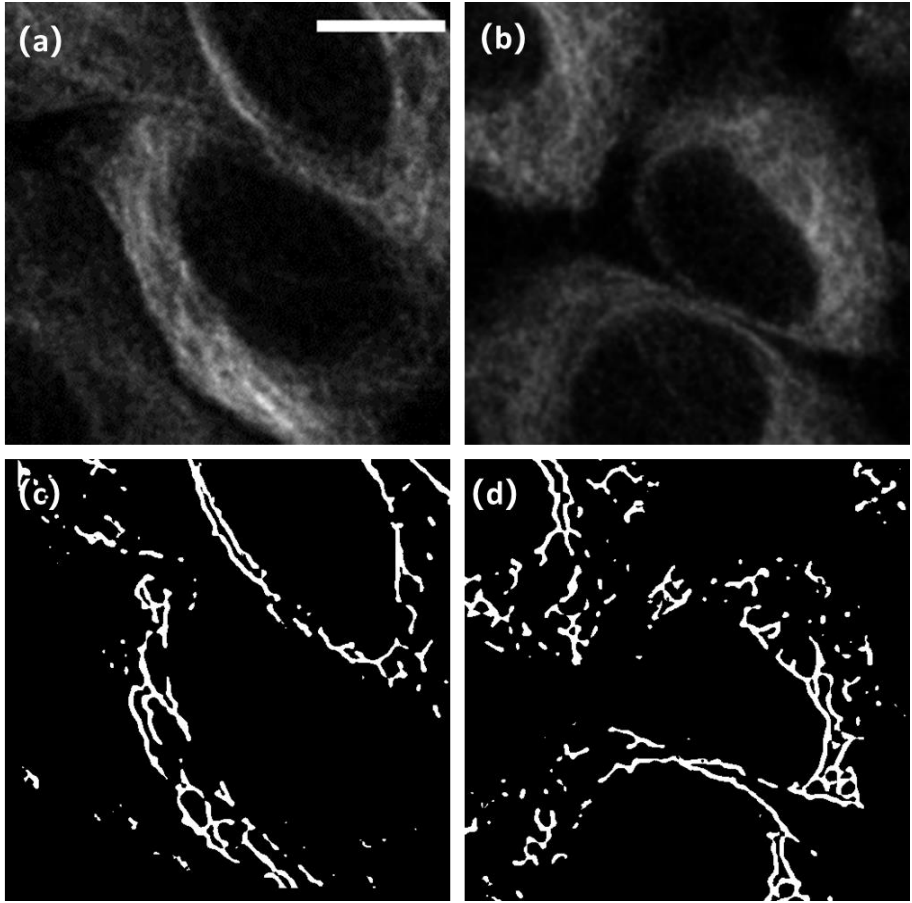

Fig. 6. A preview of SR_MUI dataset. (a) and (b) shows the sub-original images in SR_MUI dataset, (c) and (d) show the corresponding high-resolution sub-label images. The sub-image of original and the sub-image of label compose image pairs to form the training dataset. The white scale bar represents 10 µm.

*4.2 Implementation*

The numerical experiment is performed on PyTorch platform with Python language. We train and test the A-net network on a server with 10 NVidia RTX 2080TI GPUs. The epoch number is 200. The size of the mini-batch is 1. In the entire training process, the A-net network adopts the Adam optimizer. In the testing process, the test images ($2048 \times 2048$ pixels) were split into 16 sub-images of $512 \times 512$ pixel. After input the 16 sub-images into the A-net network to get the corresponding prediction images, the 16 prediction images were assembled to get the prediction image of $2048 \times 2048$ pixels. The result image is obtained after the postprocessing.

## 5. Results

Fig. 7(a) is a typical test image which has low SNR and poor resolutions. Although we can roughly distinguish the cluster of microtubule structures from the crowded backgrounds, the image cannot provide more accurate information on the distribution of microtubule. In contrast, in the result images after A-net training, noise has been significantly suppressed and the microtubule structure information is extracted from the test image. In addition, after zooming in the local image structures of both test and result images, we can see the two images (Fig. 7(c, d)) show consistent structures of microtubule.

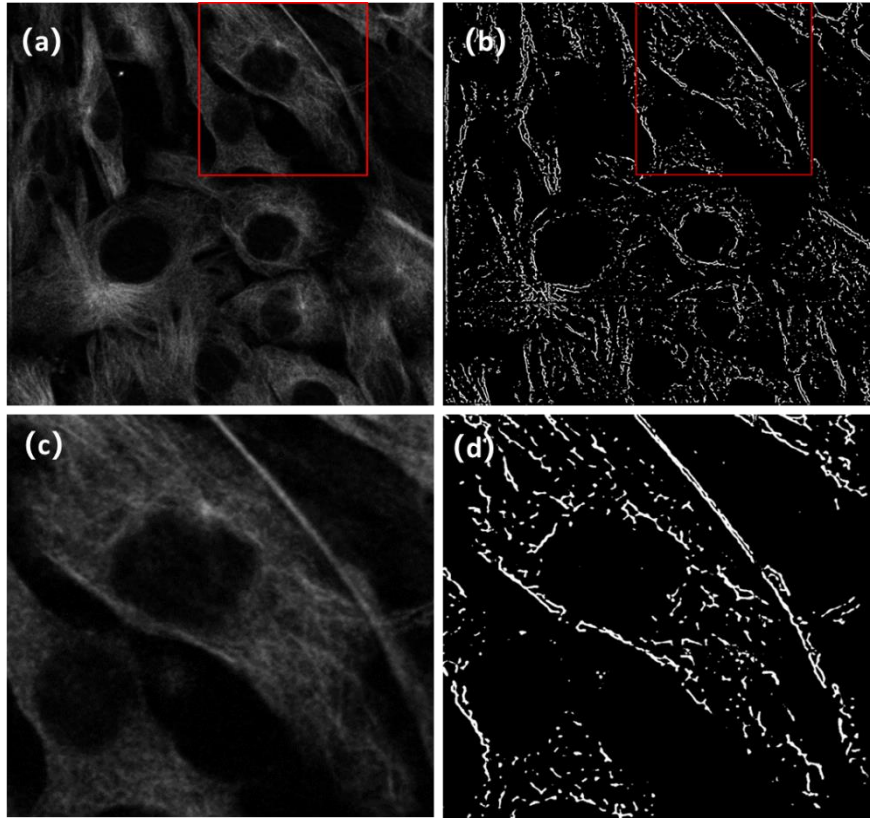

Fig. 7. Comparison between the test image and the result image. (a) A test image put into the A-net network, (b) the super-resolution result image. The test image is obtained through applying binarization on prediction images output from AU-net network and multiply the binary image with the test image. The image size is 2048×2048. (c) and (d) shows the comparison of local regions in the red boxes of the test image and the result image.

At the same time, in order to verify the consistency of microtubule structures between the test images and the result image, we overlap the test images with the result image (Fig. 8). For different test image, the microtubule structures have been concisely highlighted by the result image. The results clearly demonstrate the capability of A-net network on reserving the raw filament-like structures. It should be noted, as a result of the high noise level, some structures have been inevitably segmented, which, we believe, won't affect the overall topology of microtubule structures.

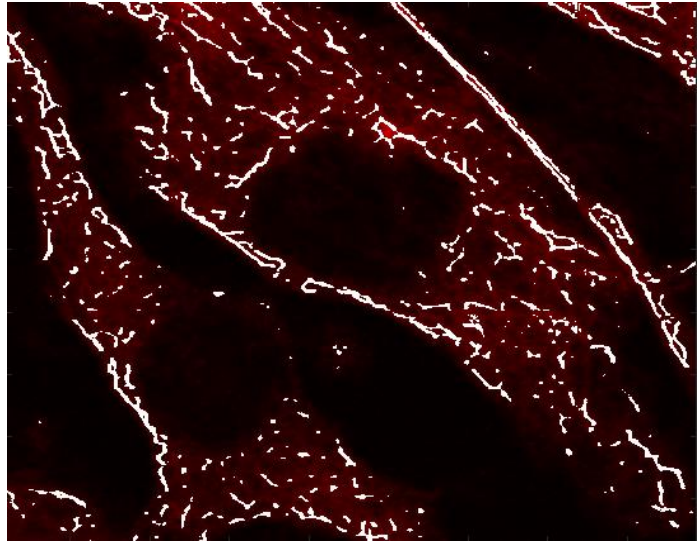

Fig. 8. Test and result images are overlapped to show the consistency of structures. The test image is shown in red and the result image is shown in white. It can be seen that the main structures in the test picture have been extracted and show high consistency with the result image.

We further compare the image intensity profiles along horizontal direction between the test and result images, as shown in Fig. 9. It's observed that the image intensity distribution of the result image has sharper peak and apparently lower noise. Overall, the result image retains a large number of structural information of the test image. We calculated the full width at half maxima (FWHM)[58] of both images. As shown in the right column of Fig. 9, the FWHM of the test image is 1.19 μm as compared to 120 nm of the result image. A super resolution with ~10 times improvement of resolution compared to the original image has been realized.

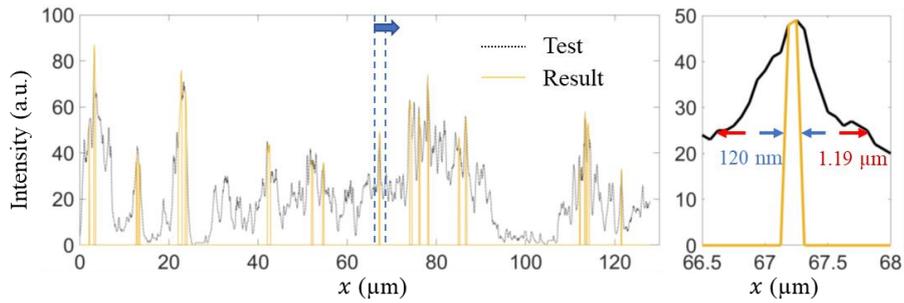

Fig. 9. Comparison of the image intensity distribution between the test image and the result image. The left figure is the intensity profiles along the horizontal direction. The right one is a zoom-in view of the left one in the marked position. The FWHM of test image is 1.19 µm, and that of the result image is 120 nm.

For the result image, the SSIM and PSNR are 0.22 and 25.88 respectively, which are surprisingly low. This is because our purpose is to extract the information of microtubule structures with super resolution, the SSIM and PSNR values cannot provide effective evaluation on image processing. Although there is currently no appropriate criterion to evaluate the processing, the improvement in visual visibility and clarity of image structure is sufficient to demonstrate the effectiveness of this algorithm.

Furthermore, we built the 3-dimensional (3D) structures of microtubule on the basis of raw and result images layer-by-layer, as shown in Fig. 10(a) and (b) respectively. Fig. 10(c) displays the 3D view of the lower left region of (b). As a result of the low signal-noise ratio of raw images, the 3D microtubule structures constructed from raw images are blur and unclear. We cannot distinguish the spatial distributions of the structures and even the skeleton. In contrast, the 3D microtubule structures constructed from result images eliminate the noise and show the structures clearly. The major biological structures are continuous which supports the effectiveness of the method.

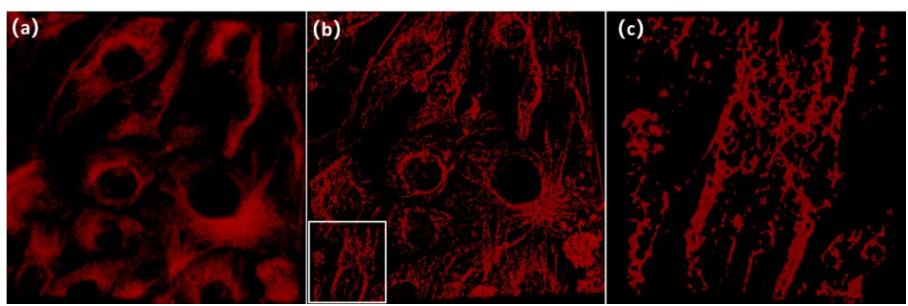

Fig. 10. 3D microtubule structures reconstructed by a set of result images. (a) 3D microtubule structures by raw images, (b) 3D microtubule structures by the result image. (c) Zoom-in view of the lower left region of (b). The influence of noise has been significantly inhibited with the cell structure clearly displayed. All the 3D microtubule structures can be found in the supplementary videos.

## 6. Conclusion

In this investigation, we proposed a new universal method based on A-net neural network and DWDC method to exact molecular structure of 3t3 fibroblast microtubule networks from poor quality confocal images, which requires relatively small data set. The experimental results indicate a 10 times improvement of spatial resolution, with a super resolution of 120 nm has been revealed from raw confocal images. The algorithm provides a general way for improving the resolution of filament-like structures and with less requirement of computation resources. The algorithm will benefit broad biological and biomedical researches, which strongly rely on optical imaging techniques.

**Acknowledgement** The investigation is supported by by National Natural Science Foundation of China (Grant No. 51927804, 61775181, 61378083).